\documentclass[reprint,aps,amsmath,amssymb,amsfonts,showkeys]{revtex4-2}
\usepackage[dvipdfmx]{graphicx}
\usepackage{dcolumn}
\usepackage{bm}
\usepackage{txfonts}
\usepackage{multirow}
\usepackage{color}

\usepackage{ulem}% Use as \sout{} for strike-out line
\begin{document}

\title{Thermoelectric transport properties of the quasi-one-dimensional dimer-Mott insulator $\beta'$-(BEDT-TTF)$_2$ICl$_2$}

\author{Kyohei~Eguchi$^1$}
\email{6222507@ed.tus.ac.jp}
\author{Takeru~Ito$^1$}
\author{Yoshiki~J.~Sato$^1$}
\thanks{Present address: Graduate School of Science and Engineering, Saitama University, Saitama 338-8570, Japan; yoshikisato@mail.saitama-u.ac.jp}
\author{Ryuji~Okazaki$^1$}
\email{okazaki@rs.tus.ac.jp}
\author{Hiromi~Taniguchi$^2$}
\affiliation{$^1$Department of Physics and Astronomy, Tokyo University of Science, Noda 278-8510, Japan\\
$^2$Graduate School of Science and Engineering, Saitama University, Saitama 338-8570, Japan}

\begin{abstract}

Low-dimensional materials, in which the electronic and transport properties are drastically modified in comparison to those of three-dimensional bulk materials, yield a key class of thermoelectric materials with high conversion efficiency.
Among such materials, the organic compounds may serve peculiar properties owing to their unique molecular-based low-dimensional structures
with highly anisotropic molecular orbitals.
Here we present the thermoelectric transport properties of the quasi-one-dimensional dimer-Mott insulator $\beta'$-(BEDT-TTF)$_2$ICl$_2$,
where BEDT-TTF stands for bis(ethylenedithio)-tetrathiafulvalene. 
We find that the thermopower exhibits typical activation-type temperature variation expected for insulators but its absolute value is anomalously large compared to the expected value from the activation-type temperature dependence of the electrical resistivity.
Successively, the Jonker-plot analysis, in which the thermopower is usually scaled by the logarithm of the resistivity, shows an unusual relation among such transport quantities.
We discuss a role of the low dimensionality for the enhanced thermopower along with recent observations of such a large thermopower in several low-dimensional materials.

\end{abstract}

\maketitle

\section{introduction}

The thermoelectric effect is a mutual conversion between electrical and heat flows, which serves an environmentally friendly energy harvesting technology that enables to 
utilize a large amount of waste heat \cite{Hebert2016,He2017,Shi2020,Yan2022}.
The low dimensionality is one of the key concepts in enhancing the thermoelectric conversion
efficiency \cite{Hicks1993} as is examined in various forms of materials such as 
one-dimensional (1D) nanowires \cite{Hochbaum2008,Boukai2008} and two-dimensional (2D) thin films of superlattices \cite{Venkatasubramanian2001,Ohta2007,Shimizu2016}.
Indeed, the low dimensionality significantly affects the transport properties;
the electronic density of states (DOS) strongly depends on the dimensionality of the systems,
and as a consequence, the thermopower, a measure of the energy derivative of the DOS \cite{Behniabook}, is greatly enhanced. 
Also, the phonon thermal conductivity is drastically suppressed by introducing 
low-dimensional structures \cite{Venkatasubramanian_ph}.

The organic compounds may serve a unique class of low-dimensional thermoelectrics because they are composed of various organic molecules with highly anisotropic molecular orbitals \cite{Lindorf2020,Eryilmaz2023}.
The electronic band structure and the resultant Fermi surfaces of such organic materials are 
indeed low dimensional, as is investigated experimentally and theoretically \cite{Jerome1982,Miyazaki2003,Kobayashi2004,Kartsovnik2004,Seo2006,Mori2006}.
Interestingly, the low dimensional nature is significant not only in the 
metallic state but also in the insulating phase such as charge-ordered 
insulators.
This feature is different from other low-dimensional systems with an
insulating nature;
for example, in the 2D oxide Mott insulators such as cuprates \cite{Komiya2002} and ruthenates \cite{Nakamura2002}, the transport anisotropy defined as $\gamma = \rho_{\rm out}/\rho_{\rm in}$, where 
$\rho_{\rm in}$~$(\rho_{\rm out})$ is the in-plane (out-of-plane) resistivity,
is not so large compared to that of two-dimensional metal with cylindrical Fermi surfaces.
On the contrary,
the resistivity anisotropy of the 2D organic compounds with 
the charge-ordered or charge-glass insulating phase is unexpectedly large as $\gamma \sim 10^3$ \cite{Sato2020},
indicating that the low dimensionality is significant in such organic systems,
probably owing to the anisotropically extended molecular orbitals.

Such a large transport anisotropy in the organic materials thus poses the question how such a low dimensionality affects the thermoelectric effect.
Indeed, the anomalously large thermopower has been observed in several low-dimensional 
organic systems.
In the low-dimensional organic molecular conductor $\beta'$-(BEDT-TTF)$_2$AuCl$_2$ [BEDT-TTF (ET) = bis(ethylenedithio)-tetrathiafulvalene],
the resistivity exhibits a typical insulating temperature dependence of $\rho(T) \propto \exp(E_{\rm g}/2k_{\rm B}T)$,
where $k_{\rm B}$, $E_{\rm g}$, and $T$ are the Boltzmann constant, the band-gap energy, and the temperature, respectively,
and $E_{\rm g}$ is estimated as $E_{\rm g} \approx 230$ meV \cite{Taniguchi2005}.
According to the conventional formula of the thermopower for insulators, 
the value of the thermopower is roughly estimated as 
$S = E_{\rm g}/2eT \approx 580$~$\mu$V/K
at $T=200$~K ($e$ being the elementary charge) from the above band-gap energy.
In contrast,
the experimental value exceeds 1~mV/K at $T=200$~K \cite{Kiyota2018}.
In addition, the 2D organic salt $\alpha$-(ET)$_2$I$_3$ 
also exhibits large thermopower in the charge-ordered insulating phase \cite{Kouda2022},
implying an unusual mechanism to enhance the thermopower beyond the conventional thermopower formula.

In the present study,
we focus on the related quasi-one-dimensional (q-1D) dimer-Mott insulator
$\beta'$-(ET)$_2$ICl$_2$ \cite{Kobayashi1986,Emge1986},
the crystal structure of which is
composed of ICl$_2$ and ET layers stacked alternately
along the $a^*$ axis [Fig. 1(a)].
The ET molecules are strongly dimerized [Fig. 1(b)] and stacked along the $b$ axis to form a conducting square lattice in the $bc$ plane [Fig. 1(c)]. 
The optical property is q-1D \cite{Kuroda1988,Hashimoto2015} while the 2D transport behavior has also been reported \cite{Tajima2008}.
The insulating layer consists of the monovalent anion ICl$_2^{-}$, and the hole transferred to the ET plane is localized on the dimer to act as a Mott insulator (the dimer Mott insulator).
Indeed, the degree of freedom in the dimerized molecules shows interesting electronic and magnetic phenomena \cite{Iguchi2013,Yoneyama2015,Hattori2017,Sawada2018}.
We measured the resistivity and the thermopower of $\beta'$-(ET)$_2$ICl$_2$
and find that
the thermopower of this compound also becomes large as compared to the activation energy estimated from the 
temperature dependence of the resistivity.
Through the Jonker-plot analysis in various low-dimensional systems \cite{Jonker1968},
we suggest that the reduced dimensionality may be crucial for such enhanced thermopower.

\section{Methods}

Single-crystalline samples were grown using an electrochemical method \cite{Taniguchi2003,Taniguchi2006}.
We determined 
the crystal orientation by using the polarized infrared reflectivity spectra measured by using a Fourier-transform infrared (FTIR) spectrometer \cite{Kuroda1988,Hashimoto2015}. 
The resistivity and the thermopower were simultaneously measured using a conventional four-probe method and a steady-state method, respectively \cite{Yamanaka2022,Sato2023} [Fig. 1(d)]. 
The direction of the electrical and heat currents is along the $c$ axis.
A manganin-constantan differential thermocouple was attached to the sample by using a carbon paste 
to measure the thermopower.
The temperature gradient ($|\nabla T|\simeq 0.5$~K/mm) was applied by using a resistive heater
and the thermoelectric voltage from the wire leads was subtracted. 

%:fig1
\begin{figure}[t]% Use * option to use single column figure
\begin{center}
\includegraphics[width=8cm]{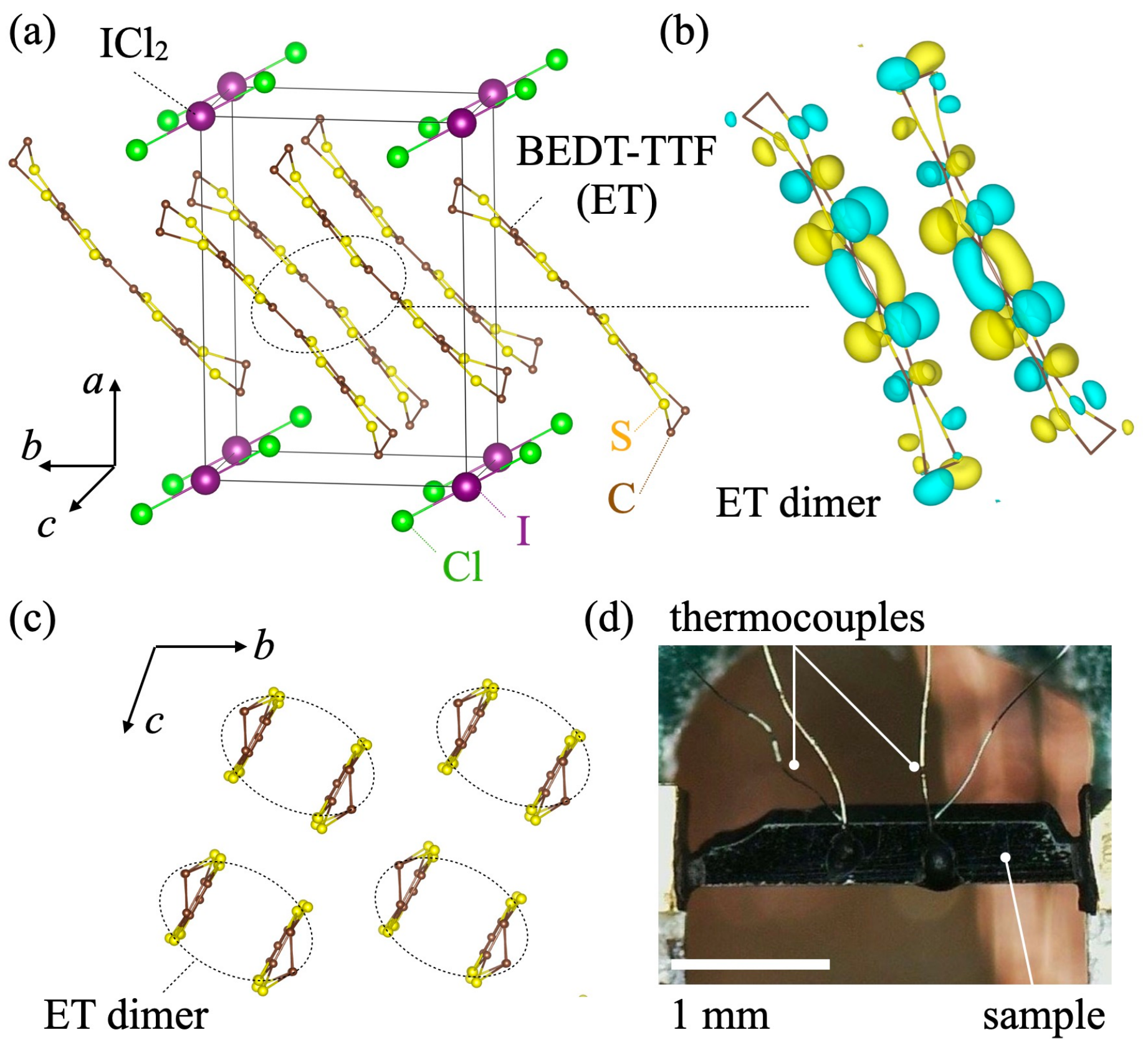}
\caption{
(a) The crystal structure of $\beta'$-(ET)$_2$ICl$_2$ drawn by VESTA \cite{Momma2011}. Hydrogen atoms are not drawn for clarity.
(b) The highest occupied molecular orbital of the dimerized ET molecules in $\beta'$-(ET)$_2$ICl$_2$.
(c) ET molecule arrangement viewed along the long axis of ET molecules. The dotted ovals represent dimerized ET molecules.
(d) Photograph of the $\beta'$-(ET)$_2$ICl$_2$ single crystal for the transport measurement.
}
\end{center}
\end{figure}

We also performed
first-principles calculations based on density functional theory (DFT) using QUANTUM ESPRESSO \cite{Giannozzi2009,Giannozzi2017,Giannozzi2020}
to obtain the molecular orbitals. 
We used the projector-augmented-wave pseudopotentials with the PerdewBurke-Ernzerhof generalized-gradient-pproximation (PBE-GGA) exchange-correlation functional. The cutoff energies for plane waves and charge densities were set to 70 and 560 Ry, respectively, and the $k$-point mesh was set to a
$10\times10\times10$ uniform grid to ensure the convergence.

\section{Results and Discussion}

%:fig1
\begin{figure}[t]% Use * option to use single column figure
\begin{center}
\includegraphics[width=9cm]{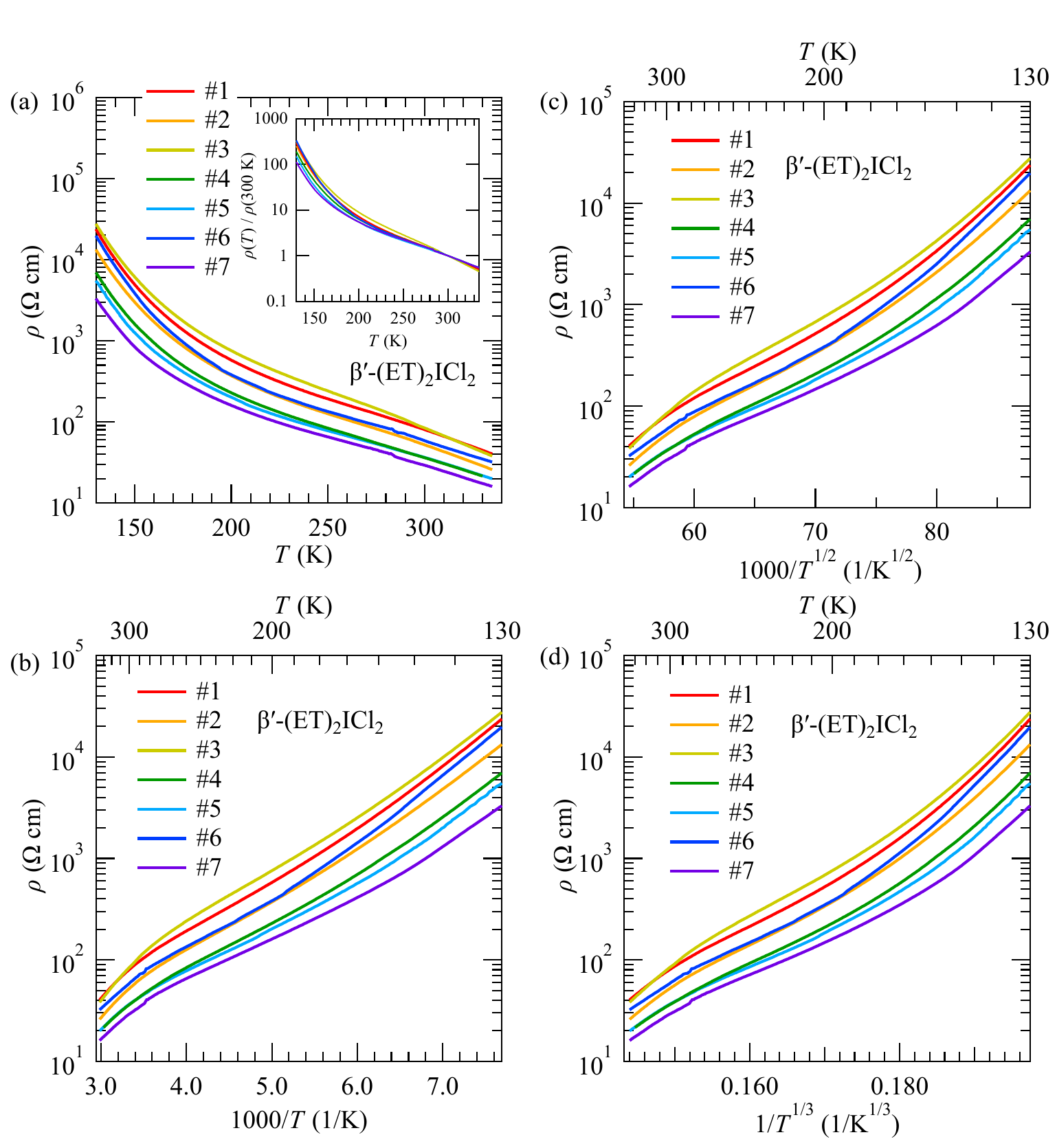}
\caption{
(a) Temperature dependence of the resistivity of $\beta'$-(ET)$_2$ICl$_2$ single crystals. The inset shows the temperature dependence of the resistivity normalized by the value at $T=300$~K.
(b) Arrhenius plot for the resistivity.
(c,d) 1D and 2D VRH plots for the resistivity.
}
\end{center}
\end{figure}

Figure 2(a) shows the temperature dependence of the electrical resistivity of $\beta'$-(ET)$_2$ICl$_2$ single crystals. 
We measured the transport properties of seven single crystals to examine the sample dependence. 
Although all of the crystals exhibit an insulating temperature dependence,
the absolute values are slightly different among the measured samples.
As shown in the inset of Fig. 2(a), 
the normalized resistivity data are also different from each other,
indicating an intrinsic sample dependence,
while the observed sample-dependent resistivity
also comes from the uncertainty of the measured sample size and the length between the voltage contacts.
The observed sample dependence will be discussed later along with the results of the thermopower experiments.
Figure 2(b) depicts the Arrhenius plot for the resistivity.
The data slightly deviate from the $1/T$ dependence probably due to the 
variable-range-hopping (VRH) conduction at low temperatures.
In the VRH model, the resistivity is predicted to follow $\rho \propto \exp\left(T_{0}/T\right)^{1/(d+1)}$, where $T_{0}$ is a constant and $d$ is the dimensionality of the system \cite{Mott1968,Efros1975}. 
Figures 2(c) and (d) depict the 1D and 2D VRH models for the resistivity, respectively. 
Because of the comparable fitting quality of these models, it is difficult to conclude whether 1D or 2D VRH is a better fit for the experimental data.
This point is also discussed in the following section.

%:fig2
\begin{figure}[t]
\begin{center}
\includegraphics[width=7.5cm]{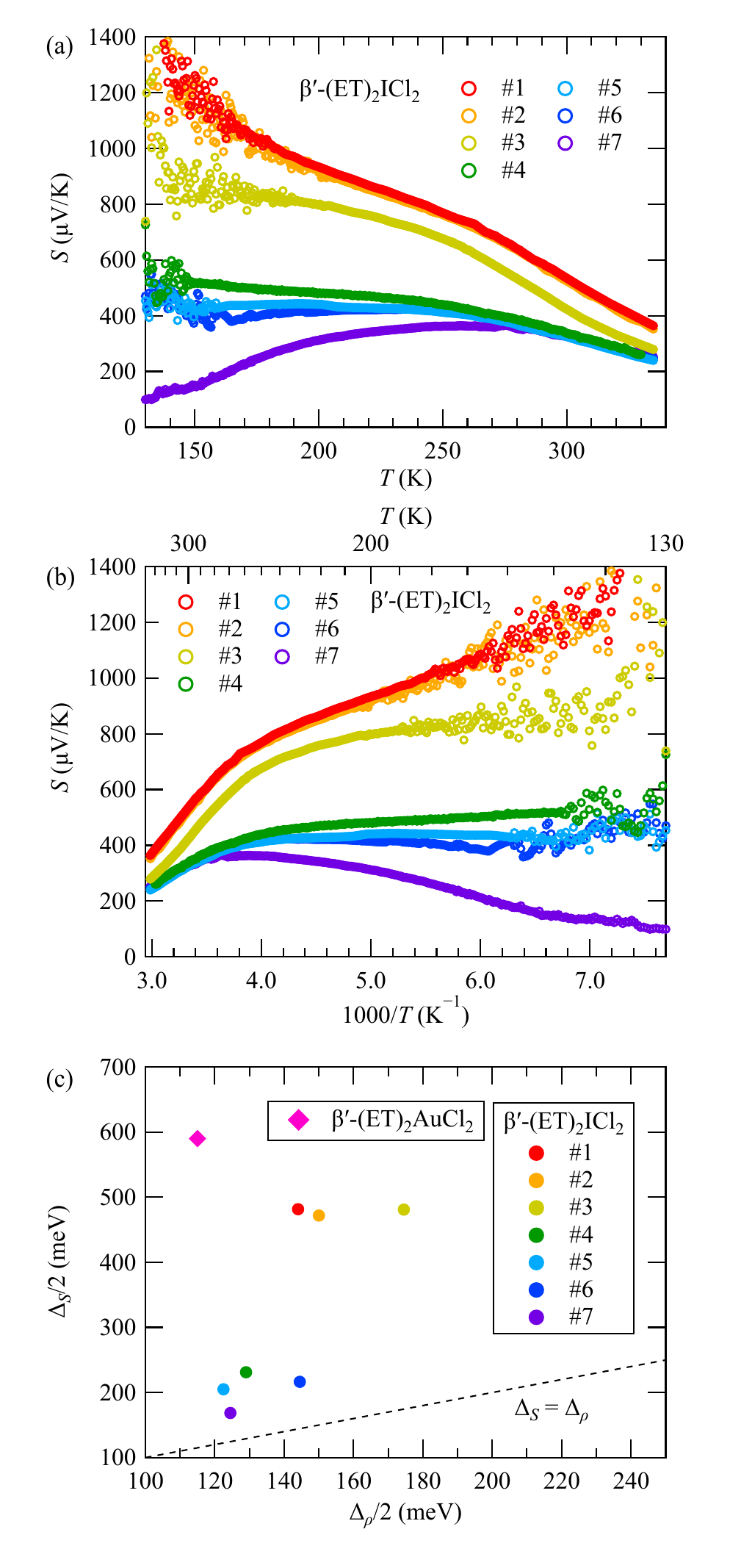}
\caption{
(a) Temperature dependence of the thermopower.
(b) Thermopower as a function of the inverse temperature.
(c) Comparison of the activation energies, $\Delta_{S}/2$ and $\Delta_{\rho}/2$.
}
\end{center}
\end{figure}

We then discuss the thermopower of $\beta'$-(ET)$_2$ICl$_2$ single crystals. 
Figure 3(a) depicts the thermopower as a function of temperature,
in which a sample dependence is clearly observed.
Near room temperature,
the thermopower increases with cooling as seen in semiconducting materials.
On the other hand, the sample dependence on the temperature variation of the thermopower at lower temperature below 200~K is significant.
Note that the low-temperature thermopower data are scattered owing to the high resistance of the measured samples.
At the present stage the details are not clear but
we infer that the observed sample dependence may originate from the VRH conduction at low temperatures,
because the VRH thermopower exhibits an increasing function of temperature
as $S\propto T^{(d-1)/(d+1)}$ in the $d$-dimensional VRH system,
while the thermopower shows a decreasing function
as $S\propto 1/T$ in semiconductors \cite{Zvyagin1973,Burns1985,Lien1999,Takahashi2012,Machida2016,Yamamoto2022}.
In the present case,
the amount of intrinsic defects and/or crystalline imperfections may be slightly different among the samples,
and consequently, the temperature dependence of the low-temperature thermopower may vary depending on the sample.
In the related ET salts \cite{Ivek2017}, such a crystalline imperfection or disorder is related to the anion layers and strongly affects the transport properties. In addition, a previous x-ray topograph experiment reveals good crystallinity of $\beta'$-(ET)$_2$ICl$_2$ single crystals \cite{Taniguchi2006} but also shows weak streak-like patterns, indicating a slight distortion in the lattice planes.

In contrast to the strongly sample-dependent thermopower below 200~K,
the thermopower data show rather similar semiconducting behavior near room temperature, as is also seen in the $S$ vs. $1/T$ plot in Fig. 3(b).
Based on a thermopower formula of
\begin{align}
    S=\frac{k_{\rm B}}{q}\left[\frac{\Delta_S}{2k_{\rm B}T} + \left( r + \frac{5}{2}\right)\right],
\end{align}
where $q$ is the carrier charge \cite{nakajima2024}, we then evaluate the activation energy $\Delta_S/2$ from the slope in the linear fitting for the $S$ vs. $1/T$ plot.
In Eq. (1), $r$ is defined as the exponent in the energy dependence of the scattering time $\tau(\varepsilon) \propto \varepsilon^r$, and $r = -1/2$ ($3/2$) corresponds to the electron-phonon (ionized impurity) scattering.
It should be noted that the scattering mechanism and the value of $r$ remain an open question in $\beta'$-(ET)$_2$ICl$_2$ \cite{Tajima2008}, and we then focus on the slope value in the linear fitting for the $S$ vs. $1/T$ plot.

The obtained activation energy $\Delta_S$/2 is plotted in Fig. 3(c)
as a function of the activation energy $\Delta_{\rho}/2$ estimated from the resistivity data near room temperature by using
\begin{align}
    \rho=\rho_0\exp\left(\frac{\Delta_{\rho}}{2k_{\rm B}T}\right).
\end{align}
Note that we used the same fitting range between $T=280$~K and 330~K for these quantities.
Interestingly, the thermopower activation energy $\Delta_S$/2 well exceeds the resistivity one,
whereas these activation energies $\Delta_{S}/2$ and $\Delta_{\rho}/2$ are identical in an ideal case.
It should be noted that the optical gap in $\beta'$-(ET)$_2$ICl$_2$ is about 2000 $\rm{cm}^{-1}$ ($\approx 250$~meV) \cite{Kuroda1988,Hashimoto2015}, which is 
close to the resistivity band-gap energy $\Delta_{\rho}$,
indicating that the thermopower is anomalously enhanced in $\beta'$-(ET)$_2$ICl$_2$.
In Fig. 3(c), 
we also plot the relation between these two activation energies of the related dimer-Mott insulator $\beta'$-(ET)$_2$AuCl$_2$ \cite{Taniguchi2005,Kiyota2018}.
The thermopower of $\beta'$-(ET)$_2$AuCl$_2$ is also significantly large and difficult to be explained by using the 
activation energy estimated from the resistivity.

%:fig4
\begin{figure}[t]
\begin{center}
\includegraphics[width=8.5cm]{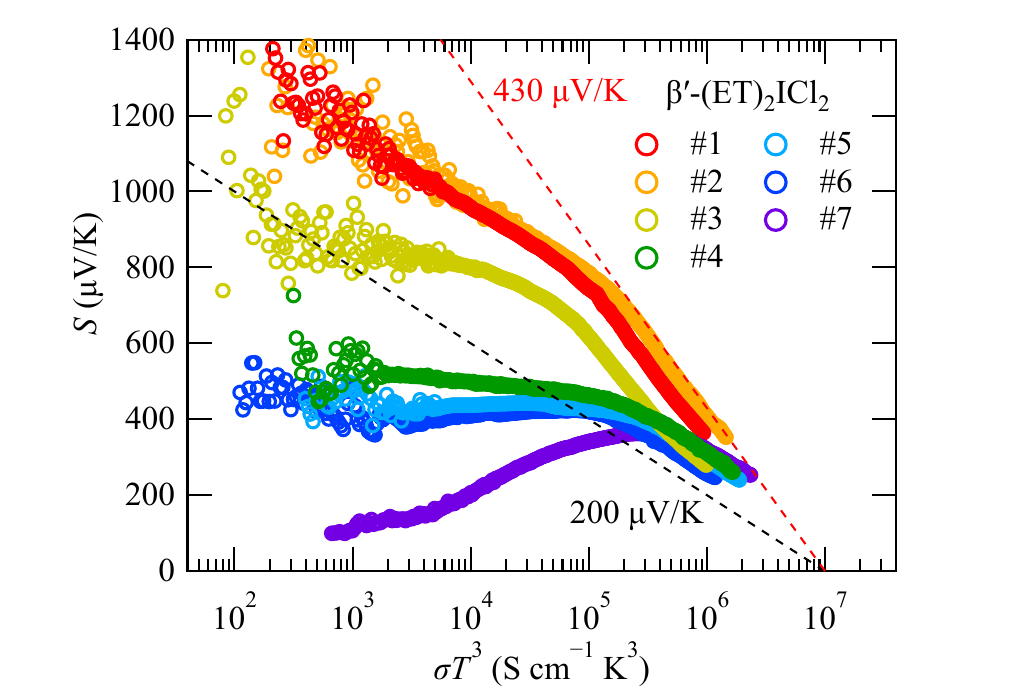}
\caption{
The Jonker plot for the measured samples of $\beta'$-(ET)$_2$ICl$_2$.
The thermopower is plotted as a function of $\sigma T^3$
to adapt the temperature-dependent mobility.
}
\end{center}
\end{figure}

We then consider the Jonker-plot analysis. 
Based on the temperature dependence of the carrier concentration $n$ in the 
activation regime of
\begin{align}
    n=n_0\exp\left(-\frac{E_{\rm g}}{2k_{\rm B}T}\right),
    \label{nnn}
\end{align}
where $n_0$ is a proportional coefficient, the activation-type thermopower is given as
\begin{align}
S=\frac{k_{\rm B}}{q}\left[\frac{E_{\rm g}}{2k_{\rm B}T} + \left( r + \frac{5}{2}\right)\right]
    = - \frac{k_{\rm B}}{q} \left[\ln n - \ln n_0 - \left( r + \frac{5}{2}\right) \right].
    \label{jonkn}
\end{align}
This relationship is known as a universal relation, and the slope between 
$S$ and $\ln n$ is given by the universal constant of $k_B/e \simeq 86$~$\mu$V/K.
(In the common logarithm plot, the slope is $k_B\ln 10/e \simeq 200$~$\mu$V/K.)
It should be noted that this slope value of 200~$\mu$V/K is 
the maximum value in the conventional semiconducting picture,
because bipolar electron-hole contributions usually reduce the absolute value of the thermopower.

The Jonker relation is often referred as the relation between $S$ and the conductivity $\sigma = \rho^{-1} = en\mu$ when the mobility $\mu$ exhibits relatively weak temperature dependence.
In $\beta'$-(ET)$_2$ICl$_2$, however, 
the carrier mobility shows an unusual temperature dependence of $\mu(T)\propto T^{-3.5}$,
the exponent of which is distinct from that of conventional one expected from
the impurity and/or the phonon scattering \cite{Tajima2008}.
The activated population in Eq. (\ref{nnn}) is then given as
\begin{align}
    \exp\left(-\frac{E_{\rm g}}{2k_{\rm B}T}\right)=
    \frac{n}{n_0}=\frac{\sigma}{e\mu n_0}
    = \frac{\sigma T^3}{A},
\end{align}
where $A$ is a constant. Here, in addition to the above $T^{-3.5}$ dependence of the mobility,
we also consider the weak temperature dependence of $n_0 \propto T^{0.5}$ for the 1D case \cite{Ashcroft}, since the optical conductivity shows the 1D characteristic \cite{Kuroda1988,Hashimoto2015}.
Note that the results are almost the same qualitatively when we use the temperature dependence of $n_0 \propto T$ for the 2D case.
Therefore, the Jonker relation yields
\begin{align}
    S 
    = - \frac{k_{\rm B}}{q} \left[\ln (\sigma T^3) - \ln A - \left( r + \frac{5}{2}\right)\right].
    \label{jonk1}
\end{align}
Figure 4 shows the thermopower as a function of $\sigma T^3$
for the measured $\beta'$-(ET)$_2$ICl$_2$ single crystals,
which is the above Jonker-plot analysis considering the temperature-dependent mobility.
Here the slope near room temperature is essential because the Jonker analysis [Eq. (\ref{jonk1})] is based on the transport properties in semiconductors as described by Eqs. (1-3).
Although the data are sample-dependent as mentioned before and the slope is not constant,
the maximum slope values in several samples well exceed the universal constant of 200~$\mu$V/K in magnitude,
implying an unusual mechanism to enhance the thermopower in this system.
The sample dependence on the slope value of the present Jonker plot will be mentioned later.

%:fig5
\begin{figure}[t]
\begin{center}
\includegraphics[width=9cm]{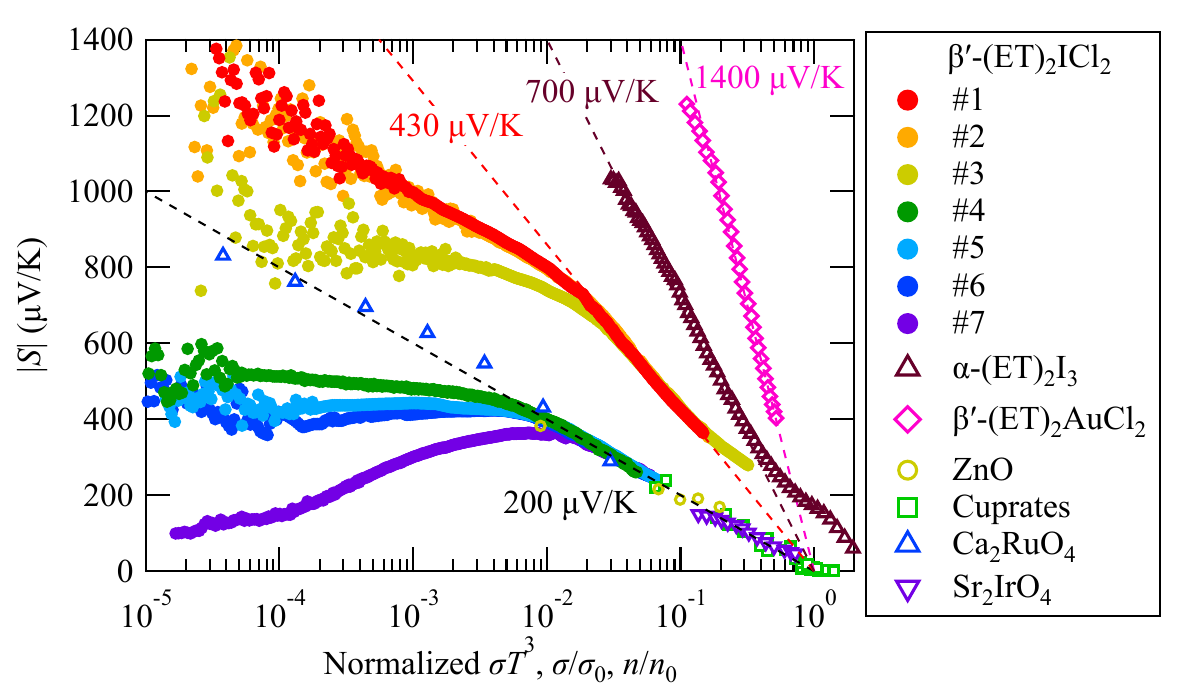}
\caption{
The Jonker plot for various low-dimensional materials.
In $\beta'$-(ET)$_2$ICl$_2$, the thermopower is shown as a function of the normalized $\sigma T^3$.
For $\beta'$-(ET)$_2$AuCl$_2$ \cite{Taniguchi2005,Kiyota2018}, $\alpha$-(ET)$_2$I$_3$ \cite{Kouda2022}, and Sr$_2$IrO$_4$ \cite{Pallecchi2016}, the horizontal axis represents the normalized conductivity $\sigma/\sigma_0$. For the cuprate superconductors \cite{Obertelli1992}, Ca$_2$RuO$_4$ \cite{Nishina2017}, and ZnO \cite{Ohtaki1996}, 
the horizontal axis represents the normalized carrier concentration $n/n_0$.
The normalization factor is the
horizontal intersect of the extrapolated slope line for each material.
}
\end{center}
\end{figure}

To examine the anomalous Jonker relation in more detail,
we compare the present data to other low-dimensional systems.
In Fig. 5, we represent the thermopower of $\beta'$-(ET)$_2$ICl$_2$ as 
a function of normalized $\sigma T^3$, in which the normalization factor was determined as the horizontal intersection of the extrapolated slope line for each sample.
We also plot the data of the organic materials $\beta'$-(ET)$_2$AuCl$_2$ \cite{Taniguchi2005,Kiyota2018} and $\alpha$-(ET)$_2$I$_3$ \cite{Kouda2022}, as well as 
the dataset of the two-dimensional oxide materials
such as the cuprates \cite{Obertelli1992}, the ruthenate Ca$_2$RuO$_4$ \cite{Nishina2017}, and the spin-orbit Mott insulator Sr$_2$IrO$_4$ \cite{Pallecchi2016}.
The data of the three-dimensional material ZnO \cite{Ohtaki1996} is also plotted as a reference.
The horizontal axis represents the normalized carrier concentration $n/n_0$
for the cuprates, Ca$_2$RuO$_4$, and ZnO.
On the other hand,
for $\beta'$-(ET)$_2$AuCl$_2$, $\alpha$-(ET)$_2$I$_3$, and Sr$_2$IrO$_4$,
the horizontal axis shows the 
normalized conductivity $\sigma/\sigma_0$ owing to the limited data, and thus 
the contributions from the temperature dependence of the mobility should be included.
Nevertheless, the observed magnitudes of the slope (430~$\mu$V/K in $\beta'$-(ET)$_2$ICl$_2$, 1400~$\mu$V/K in $\beta'$-(ET)$_2$AuCl$_2$, and 700~$\mu$V/K in $\alpha$-(ET)$_2$I$_3$)
are indeed unusual, in sharp contrast to the conventional slope value of 200~$\mu$V/K observed in the other materials including the 2D layered oxides.
Note that, although a scaling relation of $S\propto \sigma^{-1/4}$ has been proposed for the conducting polymer thermoelectrics \cite{Glaudell2015,Kang2017,Lepinoy2020}, here we adapt the Jonker analysis to discuss the thermopower of materials with low-dimensional crystal structures.

The origin of the enhanced thermopower in low-dimensional organic systems is still unclear, but we speculate that 
the highly anisotropic molecular orbitals in the organic compounds may
be important \cite{Kitou2017,Ishii2022};
in contrast to the low-dimensional oxide materials,
the low dimensionality may be significant in such organic systems 
owing to the $\pi$-stacked molecular orbitals elongated along the conduction plane or axis directions as depicted in Fig. 1(b). 
Although the conventional band picture is not appropriate in the present dimer Mott system with strong on-site Coulomb repulsion, we infer that the edge of the density of states where the carriers are thermally excited may show a steep change as a function of energy owing to the low dimensionality \cite{Behniabook}, resulting in the anomalously large slope in the Jonker plot.
The enhanced thermopower as well as the anomalously large slope value in the Jonker plot due to low dimensionality have been also discussed in the 2D superlattices and 2D electron gas realized in the field-effect transistors \cite{Ohta2007,Shimizu2016}.
Moreover, 
we infer that
the sample-dependent slope values in $\beta'$-(ET)$_2$ICl$_2$ shown in Fig. 4 and Fig. 5
originate from the sample-dependent amount of imperfections or disorder,
which may affect not only the low-temperature thermopower [Figs. 2(a) and 2(b)] but also the dimensionality for the transport properties.
In the 2D $\kappa$-type organic salt, for instance, 
the out-of-plane (interlayer) resistivity indeed depends on the sample purity \cite{Strack2005},
indicating that the amount of impurities can effectively influence dimensionality.
Such a sample-dependent effective dimensionality may possibly
lead to the sample-dependent slope values in the Jonker plot.
Also note that the uniaxial pressure experiments \cite{Steppke2017,Kawasugi2019} may be crucial to examine the effect of low dimensionality.

\section{summary}

To summarize,
we have measured the low-temperature thermoelectric transport properties of 
quasi-one-dimensional dimer-Mott insulator $\beta'$-(ET)$_2$ICl$_2$
single crystals.
We find a prominent enhancement of the thermopower,
which is not scaled in the conventional Jonker-plot analysis.
Interestingly, such a large thermopower has also been observed in
the related low-dimensional organic systems,
implying that the low dimensionality is a key property to realize the efficient 
organic-based thermoelectrics.

\section*{Acknowledgments}

We appreciate Mr. Hikaru Horita and Professor Masafumi Tamura (Tokyo University of Science) for the experimental help.
This work was supported by JSPS KAKENHI Grants No. 22H01166 and No. 24K06945.

%\bibliography{ref}
%apsrev4-2.bst 2019-01-14 (MD) hand-edited version of apsrev4-1.bst
%Control: key (0)
%Control: author (8) initials jnrlst
%Control: editor formatted (1) identically to author
%Control: production of article title (0) allowed
%Control: page (0) single
%Control: year (1) truncated
%Control: production of eprint (0) enabled
%

\end{document}